\documentclass[aps,reprint,superscriptaddress,floatfix,showpacs,pra]{revtex4-2}
\usepackage{graphicx}
\usepackage{amsmath}
\usepackage{amssymb}
\usepackage{amsfonts}
\usepackage{appendix}

\usepackage[unicode=true,breaklinks=true,colorlinks=true]{hyperref}
\hypersetup{citecolor=blue,urlcolor=blue}
\usepackage{newtxtext}
\usepackage{newtxmath}
\usepackage[all]{hypcap}
\usepackage{multirow}
\usepackage{booktabs}
\usepackage{bbold}
\usepackage{bm}
\usepackage{bbm}
\usepackage{tikz}
\usetikzlibrary{matrix,calc}
\usepackage{siunitx}

\newcommand{\ket}[1]{|{#1}\rangle}

\usepackage[dvipsnames]{xcolor}

\usepackage{indentfirst}
\usepackage{derivative}
\usepackage[dvipsnames]{xcolor}

\newenvironment{revisionn}{\color{Red}}{}

\newcommand{\DeleteNext}[1]{}

\begin{document}\sloppy
	
\title{Efficient time-evolution of matrix product states using average Hamiltonians}

\author{Belal Abouraya}
\affiliation{
Laboratory of Micro-optics, Faculty of Information Engineering and Technology, German University in Cairo (GUC), Egypt
}
\affiliation{Institute for Quantum Optics \& Center for Integrated Quantum Science and 
Technology, Universit\"{a}t Ulm, 89081 Ulm, Germany}

\author{Jirawat Saiphet}
\affiliation{
Institut f\"ur Theoretische Physik, Eberhard Karls Universit\"at T\"ubingen, 72076 T\"ubingen, Germany
}

\author{Fedor Jelezko}
\affiliation{Institute for Quantum Optics \& Center for Integrated Quantum Science and 
Technology, Universit\"{a}t Ulm, 89081 Ulm, Germany}

\author{Ressa S. Said}
\email{ressa.said@uni-ulm.de}
\affiliation{Institute for Quantum Optics \& Center for Integrated Quantum Science and 
Technology, Universit\"{a}t Ulm, 89081 Ulm, Germany}

\begin{abstract}
Simulating quantum many-body systems~(QMBS) is one of the long-standing, highly non-trivial challenges in condensed matter physics and quantum information due to the exponentially growing size of the system's Hilbert space. To date, tensor networks have been an essential tool for studying such quantum systems, owing to their ability to efficiently capture the entanglement properties of the systems they represent. One of the well-known tensor network architectures, namely matrix product states~(MPS), is the standard method for simulating one-dimensional QMBS. Here, we propose a simple, yet efficient, method to augment the already available MPS algorithms to simulate the dynamics of time-dependent Hamiltonians with better accuracy and a faster convergence rate, giving a second-order convergence compared to the first-order convergence of the standard method. We apply our proposed method to simulate the dynamics of a chain of single spins associated with nitrogen-vacancy color centers in diamonds, which has potential applications for practical and scalable quantum technologies, and find that our method improves the average error for a system of few NV centers by a factor of about 1000 for moderate step sizes. Our work paves the way for efficient simulation of QMBS under the influence of time-dependent Hamiltonians.
\end{abstract}

\maketitle

{\em Introduction.--}
Since the inception of the density matrix renormalization group~(DMRG)~\cite{tn4}, and its subsequent interpretation as the variational optimization of matrix product states~(MPS)~\cite{tn1}, tensor networks~(TN) have become a key element in studying quantum many-body systems~(QMBS), on account of their efficiency, accuracy, and scalability. Despite the fact that developments of TN for two or more dimensions are still elusive, they offer a wide range of well-developed methods for one-dimensional~(1-D) and quasi-1-D systems. Many TN representations exist, for instance multiscale entanglement renormalization ansatz~(MERA), matrix product states (MPS), projected entangled pair states (PEPS), and tree tensor networks (TTN). Up to now, the MPS, also known as~\textit{the tensor train decomposition}, is the most well-established representation~\cite{tn1, tn2, tn3, tn4, tn5}, with a variety of methods for simulating QMBS under time-independent Hamiltonians, such as the time-evolving block decimation~(TEBD) and MPO~$W^{II}$~\cite{tn1, tn2, tn3, tn5}.

However, the principal challenge in investigating QMBS numerically is the dynamical simulation governed by time-dependent Hamiltonians that describe many physical phenomena, such as those typically found in quantum control and spectroscopy using external time-dependent driving fields~\cite{nv4, nv5}, in Floquet Hamiltonian engineering of many-body systems~\cite{floquet}, and in open quantum system dynamics approximated by time-dependent effective Hamiltonians~\cite{opensys}.

\begin{figure}[!t]
\centering
         \includegraphics[width=0.65\linewidth]{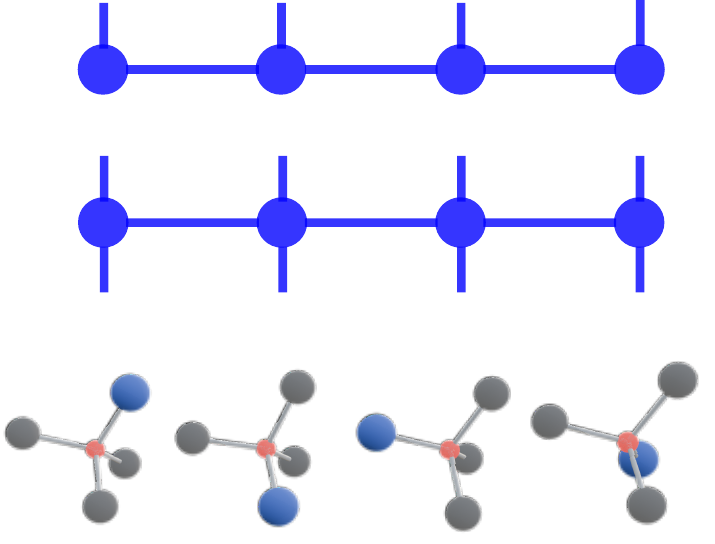}
\put(-170,110){(a)}
\put(-170,75){(b)}
\put(-170,35){(c)}
\caption{(a)~One-dimensional tensor networks of a quantum state, and (b)~the corresponding Hamiltonian representation. (c)~A chain of nitrogen-vacancy color centers in diamond. The blue and red spheres represent a pair of nitrogen and vacancy~(NV), tetrahedrally coordinated in the carbon lattice having an axial trigonal $C_{3V}$ symmetry. The negatively-charged~NV~(denoted simply as the~$NV$ throughout the paper), has an extra electron located at the vacancy site forming a spin~$(S=1)$ pair with one of the vacancy electrons, and therefore inducing spin triplet ground states. The spin chain is due to nearest-neighbor interactions of dipolar coupling (not shown in the figure), where each spin can have one out of four different quantization axes.}
\label{fig1}
\end{figure}

Our work presented in this letter is to address the above-mentioned challenge and motivated by the need for a simple-yet-highly efficient way to improve the existing MPS time-evolution solvers, or~\textit{time-steppers}, and to maintain their higher-order accuracies. 
We substitute the instantaneous Hamiltonian used inside the fixed-order MPS time-steppers by a carefully chosen average Hamiltonian, which is computed from a high-order quadrature rule. This approach leverages classical numerical integration ideas based on Simpson's rules to better approximate the time evolution operator over a short time interval. It is analytically simple and directly applicable to the existing MPS time evolution solvers, for instances the TEBD. Therefore, our method requires only a minimal change to the TN's standard numerical libraries.
Our numerical simulations demonstrate the methods' improved performance, where the improvements of up~200 and~4000 in the average error are observed in simulating systems of few spins associated with negatively-charged nitrogen-vacancy~(NV) color centers in diamond~(see~Fig.~\ref{fig1}). The NV~systems are chosen as our simulation platform owing primarily to their remarkable physical properties that  are promising for various quantum technology applications operating at ambient temperature, such as quantum registers~\cite{nv2,PhysRevX.15.021069}, and quantum sensors~\cite{RevModPhys.89.035002,PhysRevB.109.224107}. For a larger system size of NVs, our method shows a steady improvement in the average error by a factor of approximately~50.

{\em The time-evolution.--}
Simulating the dynamics of a quantum system governed by the time-dependent Schr\"odinger equation comes down to computing the time-evolution of the quantum state~$\ket{\psi}$, following
\begin{equation}
    \ket{\psi(\Delta t)} = U(\Delta t) \ket{\psi(0)},
\end{equation}
for a sufficiently small time-step~$\Delta t$, where the evolution operator~$U(\Delta t)$ is given by 
\begin{equation}
    U(\Delta t) = \exp\!\left( -i\int_0^{\Delta t} H(\tau)d\tau \right).
\end{equation}
For MPS, computing the time-evolution can be divided into two broad classes. The first class approximates the effect of the evolution operator directly on the MPS, without computing the operator itself. Meanwhile, the second class approximates the evolution operator firstly, then efficiently applies it to the MPS~\cite{tn3}. Here, we are mainly concerned with the second class of methods by focusing on the second-order TEBD and second-order complex time-step~MPO~$W^{\text{II}}$ algorithms~\cite{tebd,mpowii}. These methods are uniquely characterized as they can approximate the evolution operator for a constant Hamiltonian with a third-order error for a single-step, giving a second-order error for multiple-steps in the constructed unitary,
\begin{equation}
    U^{\left\{ \text{TEBD}, \text{II} \right\} }  = U(\Delta t) + O(\Delta t^3),
\end{equation}
for the TEBD and~MPO~$W^{\text{II}}$ unitary, respectively. Nonetheless, the operator construction is valid only for a time-independent Hamiltonian. For the time-dependent case, one can use a simple time-stepper, {\em the Riemann stepper}, which has been available in the standard TN libraries, such as~TenPy~\cite{tenpy}. It works by substituting the value of the Hamiltonian in the left end-point of the time interval into the time evolution method of choice. Starting with 
\begin{equation}
    H_{\text{Riem}}(\Delta t) \coloneq H(0),
\end{equation}
for the case of TEBD, one simply obtains  
\begin{equation}
    U^{\text{TEBD}}_{\text{Riem}}(\Delta t) = U^{\text{TEBD}}(\Delta t, H_{\text{Riem}}(\Delta t)) = U(\Delta t) + O(\Delta t^2).
\end{equation}
Alternatively, one may consider substituting an average Hamiltonian into the algorithm construction to obtain the new evolution operator. Here, we propose the average Hamiltonian in the form of
\begin{equation}
    H_{\text{Simp}}(\Delta t) \coloneq \frac{1}{6} \left( H(0) + 4 H ( \Delta t/2 ) + H(\Delta t) \right),
\end{equation}
that makes use of Simpson's integration rule to construct a new time-evolution operator that retains the same error bound. Hence, for TEBD, the new constructed operator is 
\begin{equation}
    U^{\text{TEBD}}_{\text{Simp}}(\Delta t) = U^{\text{TEBD}}(\Delta t, H_{\text{Simp}}(\Delta t)) = U(\Delta t) + O(\Delta t^3).
\end{equation}

To examine the validity of the error bounds, we provide the proof for the generic error bound that is subsequently used for the specific integrators. 
Here, we use~$\lVert .\rVert$, to denote the H\"older-$\infty$ norm of a matrix, and assume two matrices~$A(\Delta t)$, and~$\delta$, having~$\lVert \delta \rVert = O(\Delta t^k)$. Following~\cite{mat}, we obtain 
\begin{equation}
    e^{A(\Delta t) + \delta} = e^{A(\Delta t)} + \int_0^1 e^{(1-\tau)A(\Delta t)}\delta e^{\tau (A(\Delta t) + \delta)} d\tau,
\end{equation}
and hence arrive at
\begin{equation}    
\begin{aligned}
    \lVert e^{A(\Delta t) + \delta} - e^{A(\Delta t)} \rVert & = \left\lVert \int_0^1 e^{(1-\tau)A(\Delta t)}\delta e^{\tau (A(\Delta t) + \delta)} d\tau \right\rVert \\
    & \le \int_0^1 \left\lVert e^{(1-\tau)A(\Delta t)}\delta e^{\tau (A(\Delta t) + \delta)} \right\rVert d\tau \\
    & \le \int_0^1 \lVert e^{(1-\tau)A(\Delta t)} \rVert \lVert \delta \rVert \lVert e^{\tau (A(\Delta t) + \delta)} \rVert d\tau,
\end{aligned}
\end{equation}
and
\begin{equation}
\begin{aligned}
    \lVert e^{A(\Delta t) + \delta} - e^{A(\Delta t)} \rVert & \le \int_0^1 e^{(1-\tau)\lVert A(\Delta t)\rVert} \lVert \delta \rVert e^{\lVert \tau(A(\Delta t) + \delta) \rVert} d\tau \\
    & \le \int_0^1 e^{(1-\tau)\lVert A(\Delta t)\rVert} \lVert \delta \rVert e^{\lVert \tau A(\Delta t) \rVert + \lVert \tau \delta \rVert} d\tau \\
    & = \int_0^1 e^{\lVert A(\Delta t)\rVert} \lVert \delta \rVert e^{\lVert \tau \delta \rVert} d\tau. \\
\end{aligned}
\end{equation}
Assuming~$\lVert A(t) \rVert$, is bounded by~$M$, for~$t \ge 0$, we now have
\begin{equation}
\begin{aligned}
    \lVert e^{A(\Delta t) + \delta} - e^{A(\Delta t)} \rVert & \le e^{M} \int_0^1 \lVert \delta \rVert (1+\tau \lVert \delta \rVert + \dots) d\tau \\
    & \le e^{M} \int_0^1 O(\Delta t^k) (1+\tau O(\Delta t^k)) d\tau \\
    & = O(\Delta t^k).
\label{eq:error}
\end{aligned}
\end{equation}

We now use the~TEBD to obtain the error bound for the two steppers. The same derivation can however be done for~$W^{II}$. For the Riemann stepper, 
\begin{align}
U^{\text{TEBD}}_{\text{Riem}}(\Delta t) 
  &= U^{\text{TEBD}}\!\left(\Delta t, H_{\text{Riem}}(\Delta t)\right) \notag \\
  &= \exp\!\left(-i H_{\text{Riem}}(\Delta t)\, \Delta t\right) 
     + O(\Delta t^3).
\end{align}
Following~\cite{numeric}
\begin{equation}
    H_{\text{Riem}}(\Delta t) \Delta t = \int_0^{\Delta t} H(\tau) d\tau + O(\Delta t ^ 2),
\end{equation}
and substituting~$A(\Delta t) =  -i\int_0^{\Delta t} H(\tau) d\tau$~into Eq.~\ref{eq:error}, we have 
\begin{align}
U^{\text{TEBD}}_{\text{Riem}}(\Delta t) 
  &= \exp\!\left(-i \int_0^{\Delta t} H(\tau)\, d\tau\right) 
     + O(\Delta t ^ 2) + O(\Delta t ^ 3) \notag \\
  &= U(\Delta t) + O(\Delta t^2).
\end{align}
Using the same reasoning with the Simpson stepper, and following~\cite{numeric}, one obtains 
\begin{equation}
    H_{\text{Simp}}(\Delta t) \Delta t :=  \int_0^{\Delta t} H(\tau) d\tau + O(\Delta t ^ 5),
\end{equation}
and can clearly show that 
\begin{align}
U^{\text{TEBD}}_{\text{Simp}}(\Delta t) 
  &= \exp\!\left(-i \int_0^{\Delta t} H(\tau)\, d\tau\right) 
     + O(\Delta t ^ 5) + O(\Delta t ^ 3) \notag \\
  &= U(\Delta t) + O(\Delta t^3).
\end{align}

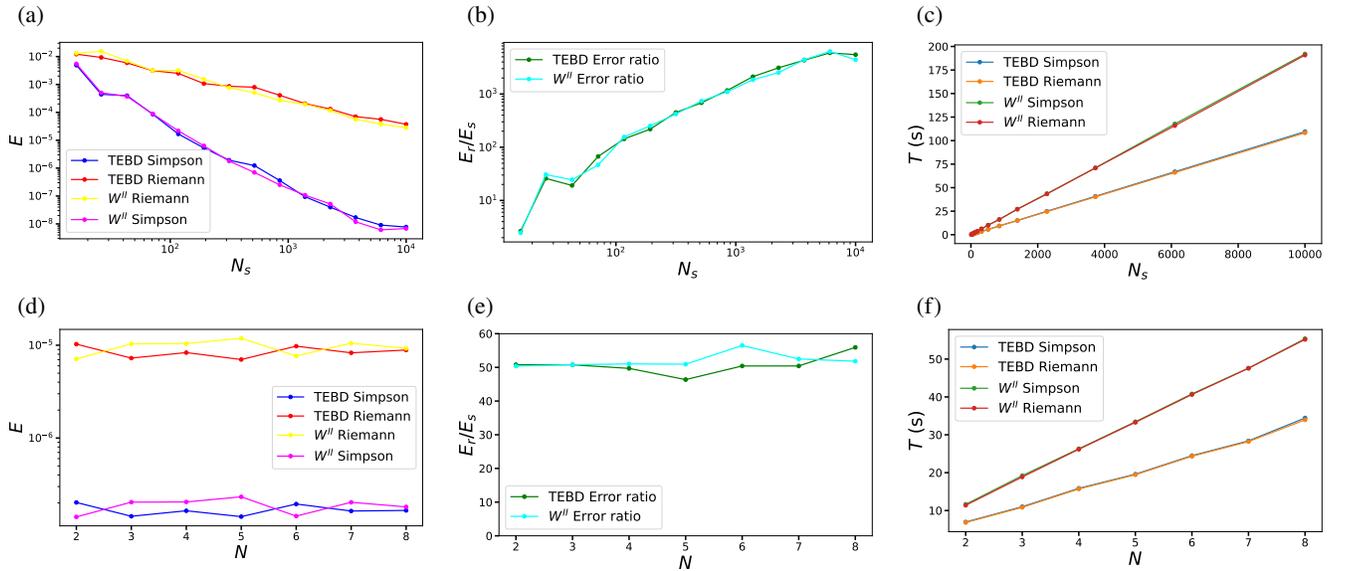
\begin{figure*}[t]
\centering
\newcommand{\imgw}{0.31\textwidth}
\newcommand{\xsep}{0.02\textwidth}
\newcommand{\ysep}{0.03\textwidth}
\newcommand{\fileA}{images/three_nv/avg_error.pdf}
\newcommand{\fileB}{images/three_nv/avg_error_ratio.pdf}
\newcommand{\fileC}{images/three_nv/avg_run_time.pdf}
\newcommand{\fileD}{images/many_nv/avg_error.pdf}
\newcommand{\fileE}{images/many_nv/avg_error_ratio.pdf}
\newcommand{\fileF}{images/many_nv/avg_run_time.pdf}
    \includegraphics[width=1.0\textwidth]{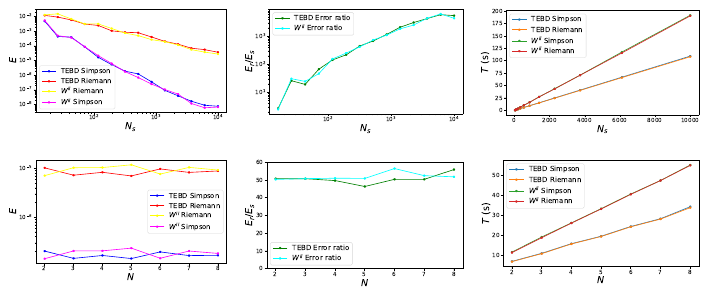} 
    	\put(-500,215){(a)}
	\put(-330,215){(b)}
	\put(-160,215){(c)}
    	\put(-500,105){(d)}
	\put(-330,105){(e)}
	\put(-160,105){(f)}
\caption{Numerical performances for various system sizes of the NV spin chain. The upper panel represents the average (a)~error, (b) error ratio, and (c) runtime of Riemann and Simpson steppers versus the number of integration steps~$N_s$, for a system of three NV centers. The lower panel amounts to the corresponding quantities, as the number of NV centers~$N$ increases: (d)~the average error, (e)~the average error ratio, and (f)~the average runtime for both steppers. The average error ratio is defined as the ratio of the average error resulting from the Riemann stepper ($E_r$) divided by the average error resulting from the Simpson stepper ($E_s)$.}
\label{fig2}
\end{figure*}

{\em Numerical simulations.--}
We numerically investigate our proposed time-evolution method to simulate a chain of $N$~spin-1~($S=1$) sites with nearest-neighbor interactions~(Fig.~\ref{fig1}), subject to an external driving pulse, representing a typical system of~$N$~NV centers driven by a microwave~(MW) pulse. 

In the frame rotating with the MW frequency and using the conventional rotating-wave approximation, the system Hamiltonian is modeled by~\cite{nv1, nv2} (here we set~$\hbar=1$, and use the unit~in~$2\pi$), as
\begin{equation}
H = H^{\text{drift}} + u(t)\cdot H^{\text{control}},
\end{equation}
where
\begin{eqnarray}
H^{\text{drift}} &=& \bigotimes_{j=1}^{N} H^0_j +H^{\text{int}}, \cr
H^0_j &=& (D - \omega_0)S_{z,j}^2 - \gamma_e B_{z,j} S_{z,j}, \cr
H^{\text{int}} &=& \sum_{j = 1}^{N-1} g_{j} S_{z,j} S_{z,j+1}, \cr
H^{\text{control}} &=& \frac{1}{2}\sum_{j = 1}^N  \cos(\zeta)S_{x,j} + \sin(\zeta)S_{y, j}',
\end{eqnarray}
with
\begin{equation}
\quad S_y' := \frac{1}{\sqrt{2}} 
\begin{pmatrix} 0 & -i & 0 \\ i & 0 & i \\ 0 & -i & 0 
\end{pmatrix}.
\end{equation}
Here, the parameter~$D$ is the NV center zero-field splitting~(ZFS) energy, while~$\gamma_e$ is the associated electronic gyromagnetic ratio of the NV spin. $B_{z,j}$ denotes the~$z$-component of the magnetic field bias experienced by the~$j$th NV center. The coupling strength between neighboring NV centers is denoted by~$g_j$. The MW driving has the carrier frequency of~$\omega_0$, and the starting phase of~$\zeta$. 

We evaluate and compare the two steppers discussed earlier in two aspects, the average error and the average runtime, as the number of steps is increased.
The averages are taken over ten random pulses, in the form of 
\begin{equation}
    u(t) = c_1 \sin( \omega_1t) + c_2 \cos(\omega_2t),
\label{eq:rand_pulse}
\end{equation}
where the $c_i$'s are randomly chosen from~$[0,5]$, and the~$\omega_i$'s are random frequencies in~$[0, 10\pi]$. The particular form of the test pulses is mainly due to the typical slow-varying modulation form of the optimal control pulses~\cite{nv5, ROSSIGNOLO2023108782, lim2024efficiencyoptimalcontrolnoisy}. The system dynamics is numerically evolved for the total time~of 0.3~$\mu$s under the application of the MW pulse. The numerical error is determined by the H\"older-$\infty$ norm~\cite{mat}, quantifying the difference between the simulation results performed by TN and the state vector methods. The state vector method used here exploits the Schr\"odinger equation numerical solver, namely{\em~\mbox{sesolve}}, in the Python-based numerical library~QuTiP~\cite{qutip}. In our simulations, the maximum absolute and relative errors are set to be~$10^{-12}$ and~$10^{-10}$, respectively to justify fairly accurate error estimations. 
We list all of the parameters used in the numerical simulations in the Table~\ref{tab:1}.

\begin{table}[!t]
    \centering
    \begin{tabular}{ll ll}
        \toprule
        Parameter & Value & Parameter & Value \\
        \midrule
        $g^{(3)}_{i}$ & 53 KHz & $g^{(N)}_{i}$ & 100 KHz \\
        $B^{(3)}_{z,1}$ & 42.82 G & $B^{(3)}_{z,2}$ & 88.31 G \\
        $B^{(3)}_{z,3}$ & 82.88 G & $B^{(N)}_{z,i}$ & 65 G \\
        $\omega^{(3)}_0$ & 2.797 GHz & $\omega^{(N)}_0$ & 2.75 GHz \\
        $\gamma_e$ & -28.025 GHz/T \cite{egyro} & $\zeta$ & 0 \\
        $D_i$ & 2.87 GHz \cite{nv3} & &\\
        \bottomrule
    \end{tabular}
    \caption{Simulation parameters. The superscript~3 and~$N$ indicates that it is used for a system of 3 and $N$ NV centers respectively. The subscript $i$ indicates that the value is the same for all NV centers. The three magnetic field parameters~$B_z$, are predeterminedly calculated to have the same resonance frequencies experimentally shown in~\cite{triplenv}, while the maximum coupling parameter~$g$ is selected. For brevity, we set~$\zeta=0$.   
    }
    \label{tab:1}
\end{table}

{\em Discussions.--}
To begin with , we tested the performance of our method as the number of steps varied by fixing the system size to three. Subsequently, we set the number of steps~$N_s$ to 1000, and varied the size of the system to evaluate the method's scalability. During the simulations, we selected the maximum bond dimension for the MPS to~3 and~16 for three and $N$ NV centers, respectively, to keep the simulation's truncation error negligibly low.

Our numerical simulation results confirm, following our expectation, that the Simpson stepper reveals improved error rates compared to that of the Riemann stepper with only a slight increase of the runtime. The such increased runtime is attributed to the increase of the number of function evaluations required by the Simpson stepper. Fig.~\ref{fig2} shows that the results for three NV centers hold for the larger system sizes with Simpson stepper yielding better error rates across all system sizes. For the case of a very simple, yet practical, pair of two NV centers, one can refer to the appendix~\ref{appendix:two_nv}.

To sum up, we introduce a simple and practical modification to commonly used MPS time-steppers for time-dependent Hamiltonians by replacing the instantaneous Hamiltonian inside a single-step propagator with a Simpson-rule average of the Hamiltonian across the step. Analytically, using this averaged Hamiltonian, our proposed method restores the native third-order local accuracy of third-order-capable integrators, such as second-order TEBD and MPO $W^{II}$.
Numerically, for the physically motivated model of driven NV centers the Simpson stepper consistently yields a lower propagation error than that of the standard Riemann substitution across the small and intermediate system sizes, while incurring only a modest increase in runtime subject to additional Hamiltonian evaluations.

{\em Outlook.--}
We conclude our letter by outlining few limitations and practical considerations. Firstly, the Simpson stepper relies on the existence and boundedness of the first four derivatives of $H(t)$, for each interval, requiring those conditions to be satisfied for the method to work. Secondly, the method increases the number of Hamiltonian evaluations per step (i.e. three per step for the Simpson method), which can dramatically increase the computational costs if the time-dependent Hamiltonian~$H(t)$ is expensive to construct.
At last, while our numerical calculations use a moderate maximum bond dimension to keep the truncation errors negligible, a problem of large-scale many-body physics, where the entanglement growth dominates significantly, is still limited by the effect of truncation errors. The Simpson stepper improves the integration error, nonetheless it does not eliminate entanglement-driven limitations. Beyond our work presented here, the Simpson averaging idea is adaptable and can be combined with a number of complementary, more efficient strategies, for instance the higher-order quadrature (e.g. Gauss–Legendre method), or multi-stage composition methods, allowing for constructing even higher-order effective Hamiltonians. Furthermore, the adaptive time-stepping procedure can be incorporated to concentrate the computational effort for the case of rapid external driving. 

\vspace{4.0mm}
\textit{Acknowledgments.} We thank Jingfu Zhang, Hendry M. Lim, and Roberto Sailer for their discussions. F.J. and R.S.S. acknowledge DFG, BMBF (CoGeQ, SPINNING, QRX), QC-4-BW, Center for Integrated Quantum Science and Technology, QTBW, ERC Synergy Grant HyperQ, and EU Projects (Spinus, C-QuENS, QCircle) for their supports.
\par
\textit{Author contributions.} B.A., J.S., and R.S.S. conceptualized the work. B.A. worked on the analytical calculations and performed numerical simulations. B.A., J.S., and R.S.S worked on the corresponding numerical data analyses. B.A. and R.S.S drafted the manuscript. F.J. and R.S.S. acquired the project funding and supervised the project. All authors read and contributed to the manuscript.
\par
\textit{Data Availability Statement.}
The datasets in this work are available from the corresponding author upon reasonable request. The processed data for the plots are publicly available~\cite{plotdata}.

\appendix

\section{Results for two NV centers}\label{appendix:two_nv}

\begin{figure*}[t]
\centering
\newcommand{\imgw}{0.31\textwidth}   
\newcommand{\xsep}{0.02\textwidth}   
\newcommand{\ysep}{0.03\textwidth}   
\includegraphics[width=1.0\textwidth]{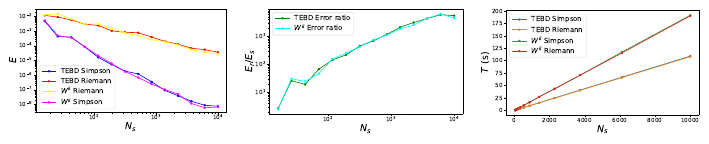}
    	\put(-500,105){(a)}
	\put(-330,105){(b)}
	\put(-160,105){(c)}
\caption{Simulation results for two NV centers. 
(a) Average error $E$ of the Riemann and Simpson steppers as a function of the number of steps $N_s$. 
(b) Average error ratio $E_r/E_s$ between the two steppers versus $N_s$. 
(c) Average runtime $T$ of both steppers as $N_s$ increases.}
\label{fig3}
\end{figure*}

The simulations for two NV centers show the similar trends observed for the other system sizes as illustrated by Fig.~\ref{fig3}, with the Simpson stepper yielding better error rates for all step sizes tested.

\begin{table}[h]
    \centering
    \label{tab:params}
    \begin{tabular}{@{} l l l l @{}}
        \toprule
        Parameter & Value & Parameter & Value \\
        \midrule
        $D_i$        & \SI{2.87}{\giga\hertz}\cite{nv3}     & $g_{1}$     & \SI{100}{\kilo\hertz} \\
        $B_{z,1}$    & $65\ \mathrm{G}$                      & $B_{z,2}$   & $20.086\ \mathrm{G}$ \\
        $\omega_0$   & \SI{2.75}{\giga\hertz}               & $\zeta$     & $0$ \\
        $\gamma_e$   & \SI{-28.025}{\giga\hertz\per\tesla}\cite{egyro} & & \\
        \bottomrule
    \end{tabular}
    \caption{Simulation parameters for the two-NV system. The subscript $i$ indicates that the value is the same for all NV centers.}
\end{table}


\begin{thebibliography}{26}%
\makeatletter
\providecommand \@ifxundefined [1]{%
 \@ifx{#1\undefined}
}%
\providecommand \@ifnum [1]{%
 \ifnum #1\expandafter \@firstoftwo
 \else \expandafter \@secondoftwo
 \fi
}%
\providecommand \@ifx [1]{%
 \ifx #1\expandafter \@firstoftwo
 \else \expandafter \@secondoftwo
 \fi
}%
\providecommand \natexlab [1]{#1}%
\providecommand \enquote  [1]{``#1''}%
\providecommand \bibnamefont  [1]{#1}%
\providecommand \bibfnamefont [1]{#1}%
\providecommand \citenamefont [1]{#1}%
\providecommand \href@noop [0]{\@secondoftwo}%
\providecommand \href [0]{\begingroup \@sanitize@url \@href}%
\providecommand \@href[1]{\@@startlink{#1}\@@href}%
\providecommand \@@href[1]{\endgroup#1\@@endlink}%
\providecommand \@sanitize@url [0]{\catcode `\\12\catcode `\$12\catcode
  `\&12\catcode `\#12\catcode `\^12\catcode `\_12\catcode `\%12\relax}%
\providecommand \@@startlink[1]{}%
\providecommand \@@endlink[0]{}%
\providecommand \url  [0]{\begingroup\@sanitize@url \@url }%
\providecommand \@url [1]{\endgroup\@href {#1}{\urlprefix }}%
\providecommand \urlprefix  [0]{URL }%
\providecommand \Eprint [0]{\href }%
\providecommand \doibase [0]{https://doi.org/}%
\providecommand \selectlanguage [0]{\@gobble}%
\providecommand \bibinfo  [0]{\@secondoftwo}%
\providecommand \bibfield  [0]{\@secondoftwo}%
\providecommand \translation [1]{[#1]}%
\providecommand \BibitemOpen [0]{}%
\providecommand \bibitemStop [0]{}%
\providecommand \bibitemNoStop [0]{.\EOS\space}%
\providecommand \EOS [0]{\spacefactor3000\relax}%
\providecommand \BibitemShut  [1]{\csname bibitem#1\endcsname}%
\let\auto@bib@innerbib\@empty
\bibitem [{\citenamefont {White}(1992)}]{tn4}%
  \BibitemOpen
  \bibfield  {author} {\bibinfo {author} {\bibfnamefont {S.~R.}\ \bibnamefont
  {White}},\ }\bibfield  {title} {\bibinfo {title} {Density matrix formulation
  for quantum renormalization groups},\ }\href@noop {} {\bibfield  {journal}
  {\bibinfo  {journal} {Physical Review Letters}\ }\textbf {\bibinfo {volume}
  {69}},\ \bibinfo {pages} {2863} (\bibinfo {year} {1992})}\BibitemShut
  {NoStop}%
\bibitem [{\citenamefont {Bridgeman}\ and\ \citenamefont {Chubb}(2017)}]{tn1}%
  \BibitemOpen
  \bibfield  {author} {\bibinfo {author} {\bibfnamefont {J.~C.}\ \bibnamefont
  {Bridgeman}}\ and\ \bibinfo {author} {\bibfnamefont {C.~T.}\ \bibnamefont
  {Chubb}},\ }\bibfield  {title} {\bibinfo {title} {Hand-waving and
  interpretive dance: an introductory course on tensor networks},\ }\href@noop
  {} {\bibfield  {journal} {\bibinfo  {journal} {Journal of Physics A:
  Mathematical and Theoretical}\ }\textbf {\bibinfo {volume} {50}},\ \bibinfo
  {pages} {223001} (\bibinfo {year} {2017})}\BibitemShut {NoStop}%
\bibitem [{\citenamefont {Orús}(2014)}]{tn2}%
  \BibitemOpen
  \bibfield  {author} {\bibinfo {author} {\bibfnamefont {R.}~\bibnamefont
  {Orús}},\ }\bibfield  {title} {\bibinfo {title} {A practical introduction to
  tensor networks: Matrix product states and projected entangled pair states},\
  }\href@noop {} {\bibfield  {journal} {\bibinfo  {journal} {Annals of
  Physics}\ }\textbf {\bibinfo {volume} {349}},\ \bibinfo {pages} {117}
  (\bibinfo {year} {2014})}\BibitemShut {NoStop}%
\bibitem [{\citenamefont {Paeckel}\ \emph {et~al.}(2019)\citenamefont
  {Paeckel}, \citenamefont {Köhler}, \citenamefont {Swoboda}, \citenamefont
  {Manmana}, \citenamefont {Schollwöck},\ and\ \citenamefont {Hubig}}]{tn3}%
  \BibitemOpen
  \bibfield  {author} {\bibinfo {author} {\bibfnamefont {S.}~\bibnamefont
  {Paeckel}}, \bibinfo {author} {\bibfnamefont {T.}~\bibnamefont {Köhler}},
  \bibinfo {author} {\bibfnamefont {A.}~\bibnamefont {Swoboda}}, \bibinfo
  {author} {\bibfnamefont {S.~R.}\ \bibnamefont {Manmana}}, \bibinfo {author}
  {\bibfnamefont {U.}~\bibnamefont {Schollwöck}},\ and\ \bibinfo {author}
  {\bibfnamefont {C.}~\bibnamefont {Hubig}},\ }\bibfield  {title} {\bibinfo
  {title} {Time-evolution methods for matrix-product states},\ }\href@noop {}
  {\bibfield  {journal} {\bibinfo  {journal} {Annals of Physics}\ }\textbf
  {\bibinfo {volume} {411}},\ \bibinfo {pages} {167998} (\bibinfo {year}
  {2019})}\BibitemShut {NoStop}%
\bibitem [{\citenamefont {Felser}(2021)}]{tn5}%
  \BibitemOpen
  \bibfield  {author} {\bibinfo {author} {\bibfnamefont {T.}~\bibnamefont
  {Felser}},\ }\emph {\bibinfo {title} {{Tree tensor networks for
  high-dimensional quantum systems and beyond}}},\ \href
  {https://doi.org/10.22028/D291-35211} {Ph.D. thesis},\ \bibinfo  {school}
  {Saarland U.} (\bibinfo {year} {2021})\BibitemShut {NoStop}%
\bibitem [{\citenamefont {Rembold}\ \emph {et~al.}(2020)\citenamefont
  {Rembold}, \citenamefont {Oshnik}, \citenamefont {M{\"u}ller}, \citenamefont
  {Montangero}, \citenamefont {Calarco},\ and\ \citenamefont {Neu}}]{nv4}%
  \BibitemOpen
  \bibfield  {author} {\bibinfo {author} {\bibfnamefont {P.}~\bibnamefont
  {Rembold}}, \bibinfo {author} {\bibfnamefont {N.}~\bibnamefont {Oshnik}},
  \bibinfo {author} {\bibfnamefont {M.~M.}\ \bibnamefont {M{\"u}ller}},
  \bibinfo {author} {\bibfnamefont {S.}~\bibnamefont {Montangero}}, \bibinfo
  {author} {\bibfnamefont {T.}~\bibnamefont {Calarco}},\ and\ \bibinfo {author}
  {\bibfnamefont {E.}~\bibnamefont {Neu}},\ }\bibfield  {title} {\bibinfo
  {title} {Introduction to quantum optimal control for quantum sensing with
  nitrogen-vacancy centers in diamond},\ }\href@noop {} {\bibfield  {journal}
  {\bibinfo  {journal} {AVS Quantum Science}\ }\textbf {\bibinfo {volume} {2}}
  (\bibinfo {year} {2020})}\BibitemShut {NoStop}%
\bibitem [{\citenamefont {M{\"u}ller}\ \emph {et~al.}(2022)\citenamefont
  {M{\"u}ller}, \citenamefont {Said}, \citenamefont {Jelezko}, \citenamefont
  {Calarco},\ and\ \citenamefont {Montangero}}]{nv5}%
  \BibitemOpen
  \bibfield  {author} {\bibinfo {author} {\bibfnamefont {M.~M.}\ \bibnamefont
  {M{\"u}ller}}, \bibinfo {author} {\bibfnamefont {R.~S.}\ \bibnamefont
  {Said}}, \bibinfo {author} {\bibfnamefont {F.}~\bibnamefont {Jelezko}},
  \bibinfo {author} {\bibfnamefont {T.}~\bibnamefont {Calarco}},\ and\ \bibinfo
  {author} {\bibfnamefont {S.}~\bibnamefont {Montangero}},\ }\bibfield  {title}
  {\bibinfo {title} {One decade of quantum optimal control in the chopped
  random basis},\ }\href@noop {} {\bibfield  {journal} {\bibinfo  {journal}
  {Reports on progress in physics}\ }\textbf {\bibinfo {volume} {85}},\
  \bibinfo {pages} {076001} (\bibinfo {year} {2022})}\BibitemShut {NoStop}%
\bibitem [{\citenamefont {Oka}\ and\ \citenamefont {Kitamura}(2019)}]{floquet}%
  \BibitemOpen
  \bibfield  {author} {\bibinfo {author} {\bibfnamefont {T.}~\bibnamefont
  {Oka}}\ and\ \bibinfo {author} {\bibfnamefont {S.}~\bibnamefont {Kitamura}},\
  }\bibfield  {title} {\bibinfo {title} {Floquet engineering of quantum
  materials},\ }\href@noop {} {\bibfield  {journal} {\bibinfo  {journal}
  {Annual Review of Condensed Matter Physics}\ }\textbf {\bibinfo {volume}
  {10}},\ \bibinfo {pages} {387} (\bibinfo {year} {2019})}\BibitemShut
  {NoStop}%
\bibitem [{\citenamefont {Santos}\ \emph {et~al.}(2024)\citenamefont {Santos},
  \citenamefont {Pereira}, \citenamefont {Abrantes}, \citenamefont {Farina},
  \citenamefont {Maia~Neto},\ and\ \citenamefont {de~Melo~e Souza}}]{opensys}%
  \BibitemOpen
  \bibfield  {author} {\bibinfo {author} {\bibfnamefont {A.~S.}\ \bibnamefont
  {Santos}}, \bibinfo {author} {\bibfnamefont {P.~H.}\ \bibnamefont {Pereira}},
  \bibinfo {author} {\bibfnamefont {P.~P.}\ \bibnamefont {Abrantes}}, \bibinfo
  {author} {\bibfnamefont {C.}~\bibnamefont {Farina}}, \bibinfo {author}
  {\bibfnamefont {P.~A.}\ \bibnamefont {Maia~Neto}},\ and\ \bibinfo {author}
  {\bibfnamefont {R.}~\bibnamefont {de~Melo~e Souza}},\ }\bibfield  {title}
  {\bibinfo {title} {Time-dependent effective hamiltonians for light--matter
  interactions},\ }\href@noop {} {\bibfield  {journal} {\bibinfo  {journal}
  {Entropy}\ }\textbf {\bibinfo {volume} {26}},\ \bibinfo {pages} {527}
  (\bibinfo {year} {2024})}\BibitemShut {NoStop}%
\bibitem [{\citenamefont {Joas}\ \emph
  {et~al.}(2025{\natexlab{a}})\citenamefont {Joas}, \citenamefont {Ferlemann},
  \citenamefont {Sailer}, \citenamefont {Vetter}, \citenamefont {Zhang},
  \citenamefont {Said}, \citenamefont {Teraji}, \citenamefont {Onoda},
  \citenamefont {Calarco}, \citenamefont {Genov}, \citenamefont {M\"uller},\
  and\ \citenamefont {Jelezko}}]{nv2}%
  \BibitemOpen
  \bibfield  {author} {\bibinfo {author} {\bibfnamefont {T.}~\bibnamefont
  {Joas}}, \bibinfo {author} {\bibfnamefont {F.}~\bibnamefont {Ferlemann}},
  \bibinfo {author} {\bibfnamefont {R.}~\bibnamefont {Sailer}}, \bibinfo
  {author} {\bibfnamefont {P.~J.}\ \bibnamefont {Vetter}}, \bibinfo {author}
  {\bibfnamefont {J.}~\bibnamefont {Zhang}}, \bibinfo {author} {\bibfnamefont
  {R.~S.}\ \bibnamefont {Said}}, \bibinfo {author} {\bibfnamefont
  {T.}~\bibnamefont {Teraji}}, \bibinfo {author} {\bibfnamefont
  {S.}~\bibnamefont {Onoda}}, \bibinfo {author} {\bibfnamefont
  {T.}~\bibnamefont {Calarco}}, \bibinfo {author} {\bibfnamefont
  {G.}~\bibnamefont {Genov}}, \bibinfo {author} {\bibfnamefont {M.~M.}\
  \bibnamefont {M\"uller}},\ and\ \bibinfo {author} {\bibfnamefont
  {F.}~\bibnamefont {Jelezko}},\ }\bibfield  {title} {\bibinfo {title}
  {High-fidelity electron spin gates for scaling diamond quantum registers},\
  }\href {https://doi.org/10.1103/PhysRevX.15.021069} {\bibfield  {journal}
  {\bibinfo  {journal} {Phys. Rev. X}\ }\textbf {\bibinfo {volume} {15}},\
  \bibinfo {pages} {021069} (\bibinfo {year} {2025}{\natexlab{a}})}\BibitemShut
  {NoStop}%
\bibitem [{\citenamefont {Joas}\ \emph
  {et~al.}(2025{\natexlab{b}})\citenamefont {Joas}, \citenamefont {Ferlemann},
  \citenamefont {Sailer}, \citenamefont {Vetter}, \citenamefont {Zhang},
  \citenamefont {Said}, \citenamefont {Teraji}, \citenamefont {Onoda},
  \citenamefont {Calarco}, \citenamefont {Genov}, \citenamefont {M\"uller},\
  and\ \citenamefont {Jelezko}}]{PhysRevX.15.021069}%
  \BibitemOpen
  \bibfield  {author} {\bibinfo {author} {\bibfnamefont {T.}~\bibnamefont
  {Joas}}, \bibinfo {author} {\bibfnamefont {F.}~\bibnamefont {Ferlemann}},
  \bibinfo {author} {\bibfnamefont {R.}~\bibnamefont {Sailer}}, \bibinfo
  {author} {\bibfnamefont {P.~J.}\ \bibnamefont {Vetter}}, \bibinfo {author}
  {\bibfnamefont {J.}~\bibnamefont {Zhang}}, \bibinfo {author} {\bibfnamefont
  {R.~S.}\ \bibnamefont {Said}}, \bibinfo {author} {\bibfnamefont
  {T.}~\bibnamefont {Teraji}}, \bibinfo {author} {\bibfnamefont
  {S.}~\bibnamefont {Onoda}}, \bibinfo {author} {\bibfnamefont
  {T.}~\bibnamefont {Calarco}}, \bibinfo {author} {\bibfnamefont
  {G.}~\bibnamefont {Genov}}, \bibinfo {author} {\bibfnamefont {M.~M.}\
  \bibnamefont {M\"uller}},\ and\ \bibinfo {author} {\bibfnamefont
  {F.}~\bibnamefont {Jelezko}},\ }\bibfield  {title} {\bibinfo {title}
  {High-fidelity electron spin gates for scaling diamond quantum registers},\
  }\href {https://doi.org/10.1103/PhysRevX.15.021069} {\bibfield  {journal}
  {\bibinfo  {journal} {Phys. Rev. X}\ }\textbf {\bibinfo {volume} {15}},\
  \bibinfo {pages} {021069} (\bibinfo {year} {2025}{\natexlab{b}})}\BibitemShut
  {NoStop}%
\bibitem [{\citenamefont {Degen}\ \emph {et~al.}(2017)\citenamefont {Degen},
  \citenamefont {Reinhard},\ and\ \citenamefont
  {Cappellaro}}]{RevModPhys.89.035002}%
  \BibitemOpen
  \bibfield  {author} {\bibinfo {author} {\bibfnamefont {C.~L.}\ \bibnamefont
  {Degen}}, \bibinfo {author} {\bibfnamefont {F.}~\bibnamefont {Reinhard}},\
  and\ \bibinfo {author} {\bibfnamefont {P.}~\bibnamefont {Cappellaro}},\
  }\bibfield  {title} {\bibinfo {title} {Quantum sensing},\ }\href
  {https://doi.org/10.1103/RevModPhys.89.035002} {\bibfield  {journal}
  {\bibinfo  {journal} {Rev. Mod. Phys.}\ }\textbf {\bibinfo {volume} {89}},\
  \bibinfo {pages} {035002} (\bibinfo {year} {2017})}\BibitemShut {NoStop}%
\bibitem [{\citenamefont {Dhungel}\ \emph {et~al.}(2024)\citenamefont
  {Dhungel}, \citenamefont {Lenz}, \citenamefont {Omar}, \citenamefont
  {Rebeirro}, \citenamefont {Luu}, \citenamefont {Younesi}, \citenamefont
  {Ulbricht}, \citenamefont {Iv\'ady}, \citenamefont {Gali}, \citenamefont
  {Wickenbrock},\ and\ \citenamefont {Budker}}]{PhysRevB.109.224107}%
  \BibitemOpen
  \bibfield  {author} {\bibinfo {author} {\bibfnamefont {O.}~\bibnamefont
  {Dhungel}}, \bibinfo {author} {\bibfnamefont {T.}~\bibnamefont {Lenz}},
  \bibinfo {author} {\bibfnamefont {M.}~\bibnamefont {Omar}}, \bibinfo {author}
  {\bibfnamefont {J.~S.}\ \bibnamefont {Rebeirro}}, \bibinfo {author}
  {\bibfnamefont {M.-T.}\ \bibnamefont {Luu}}, \bibinfo {author} {\bibfnamefont
  {A.~T.}\ \bibnamefont {Younesi}}, \bibinfo {author} {\bibfnamefont
  {R.}~\bibnamefont {Ulbricht}}, \bibinfo {author} {\bibfnamefont
  {V.}~\bibnamefont {Iv\'ady}}, \bibinfo {author} {\bibfnamefont
  {A.}~\bibnamefont {Gali}}, \bibinfo {author} {\bibfnamefont {A.}~\bibnamefont
  {Wickenbrock}},\ and\ \bibinfo {author} {\bibfnamefont {D.}~\bibnamefont
  {Budker}},\ }\bibfield  {title} {\bibinfo {title} {Near zero-field
  microwave-free magnetometry with ensembles of nitrogen-vacancy centers in
  diamond},\ }\href {https://doi.org/10.1103/PhysRevB.109.224107} {\bibfield
  {journal} {\bibinfo  {journal} {Phys. Rev. B}\ }\textbf {\bibinfo {volume}
  {109}},\ \bibinfo {pages} {224107} (\bibinfo {year} {2024})}\BibitemShut
  {NoStop}%
\bibitem [{\citenamefont {Vidal}(2004)}]{tebd}%
  \BibitemOpen
  \bibfield  {author} {\bibinfo {author} {\bibfnamefont {G.}~\bibnamefont
  {Vidal}},\ }\bibfield  {title} {\bibinfo {title} {Efficient simulation of
  one-dimensional quantum many-body systems},\ }\href
  {https://doi.org/10.1103/PhysRevLett.93.040502} {\bibfield  {journal}
  {\bibinfo  {journal} {Phys. Rev. Lett.}\ }\textbf {\bibinfo {volume} {93}},\
  \bibinfo {pages} {040502} (\bibinfo {year} {2004})}\BibitemShut {NoStop}%
\bibitem [{\citenamefont {Zaletel}\ \emph {et~al.}(2015)\citenamefont
  {Zaletel}, \citenamefont {Mong}, \citenamefont {Karrasch}, \citenamefont
  {Moore},\ and\ \citenamefont {Pollmann}}]{mpowii}%
  \BibitemOpen
  \bibfield  {author} {\bibinfo {author} {\bibfnamefont {M.~P.}\ \bibnamefont
  {Zaletel}}, \bibinfo {author} {\bibfnamefont {R.~S.}\ \bibnamefont {Mong}},
  \bibinfo {author} {\bibfnamefont {C.}~\bibnamefont {Karrasch}}, \bibinfo
  {author} {\bibfnamefont {J.~E.}\ \bibnamefont {Moore}},\ and\ \bibinfo
  {author} {\bibfnamefont {F.}~\bibnamefont {Pollmann}},\ }\bibfield  {title}
  {\bibinfo {title} {Time-evolving a matrix product state with long-ranged
  interactions},\ }\href@noop {} {\bibfield  {journal} {\bibinfo  {journal}
  {Physical Review B}\ }\textbf {\bibinfo {volume} {91}},\ \bibinfo {pages}
  {165112} (\bibinfo {year} {2015})}\BibitemShut {NoStop}%
\bibitem [{\citenamefont {Hauschild}\ \emph {et~al.}(2024)\citenamefont
  {Hauschild}, \citenamefont {Unfried}, \citenamefont {Anand}, \citenamefont
  {Andrews}, \citenamefont {Bintz}, \citenamefont {Borla}, \citenamefont
  {Divic}, \citenamefont {Drescher}, \citenamefont {Geiger}, \citenamefont
  {Hefel}, \citenamefont {Hémery}, \citenamefont {Kadow}, \citenamefont
  {Kemp}, \citenamefont {Kirchner}, \citenamefont {Liu}, \citenamefont
  {Möller}, \citenamefont {Parker}, \citenamefont {Rader}, \citenamefont
  {Romen}, \citenamefont {Scalet}, \citenamefont {Schoonderwoerd},
  \citenamefont {Schulz}, \citenamefont {Soejima}, \citenamefont {Thoma},
  \citenamefont {Wu}, \citenamefont {Zechmann}, \citenamefont {Zweng},
  \citenamefont {Mong}, \citenamefont {Zaletel},\ and\ \citenamefont
  {Pollmann}}]{tenpy}%
  \BibitemOpen
  \bibfield  {author} {\bibinfo {author} {\bibfnamefont {J.}~\bibnamefont
  {Hauschild}}, \bibinfo {author} {\bibfnamefont {J.}~\bibnamefont {Unfried}},
  \bibinfo {author} {\bibfnamefont {S.}~\bibnamefont {Anand}}, \bibinfo
  {author} {\bibfnamefont {B.}~\bibnamefont {Andrews}}, \bibinfo {author}
  {\bibfnamefont {M.}~\bibnamefont {Bintz}}, \bibinfo {author} {\bibfnamefont
  {U.}~\bibnamefont {Borla}}, \bibinfo {author} {\bibfnamefont
  {S.}~\bibnamefont {Divic}}, \bibinfo {author} {\bibfnamefont
  {M.}~\bibnamefont {Drescher}}, \bibinfo {author} {\bibfnamefont
  {J.}~\bibnamefont {Geiger}}, \bibinfo {author} {\bibfnamefont
  {M.}~\bibnamefont {Hefel}}, \bibinfo {author} {\bibfnamefont
  {K.}~\bibnamefont {Hémery}}, \bibinfo {author} {\bibfnamefont
  {W.}~\bibnamefont {Kadow}}, \bibinfo {author} {\bibfnamefont
  {J.}~\bibnamefont {Kemp}}, \bibinfo {author} {\bibfnamefont {N.}~\bibnamefont
  {Kirchner}}, \bibinfo {author} {\bibfnamefont {V.~S.}\ \bibnamefont {Liu}},
  \bibinfo {author} {\bibfnamefont {G.}~\bibnamefont {Möller}}, \bibinfo
  {author} {\bibfnamefont {D.}~\bibnamefont {Parker}}, \bibinfo {author}
  {\bibfnamefont {M.}~\bibnamefont {Rader}}, \bibinfo {author} {\bibfnamefont
  {A.}~\bibnamefont {Romen}}, \bibinfo {author} {\bibfnamefont
  {S.}~\bibnamefont {Scalet}}, \bibinfo {author} {\bibfnamefont
  {L.}~\bibnamefont {Schoonderwoerd}}, \bibinfo {author} {\bibfnamefont
  {M.}~\bibnamefont {Schulz}}, \bibinfo {author} {\bibfnamefont
  {T.}~\bibnamefont {Soejima}}, \bibinfo {author} {\bibfnamefont
  {P.}~\bibnamefont {Thoma}}, \bibinfo {author} {\bibfnamefont
  {Y.}~\bibnamefont {Wu}}, \bibinfo {author} {\bibfnamefont {P.}~\bibnamefont
  {Zechmann}}, \bibinfo {author} {\bibfnamefont {L.}~\bibnamefont {Zweng}},
  \bibinfo {author} {\bibfnamefont {R.~S.~K.}\ \bibnamefont {Mong}}, \bibinfo
  {author} {\bibfnamefont {M.~P.}\ \bibnamefont {Zaletel}},\ and\ \bibinfo
  {author} {\bibfnamefont {F.}~\bibnamefont {Pollmann}},\ }\bibfield  {title}
  {\bibinfo {title} {{Tensor network Python (TeNPy) version 1}},\ }\href
  {https://doi.org/10.21468/SciPostPhysCodeb.41} {\bibfield  {journal}
  {\bibinfo  {journal} {SciPost Phys. Codebases}\ ,\ \bibinfo {pages} {41}}
  (\bibinfo {year} {2024})}\BibitemShut {NoStop}%
\bibitem [{\citenamefont {Bernstein}(2009)}]{mat}%
  \BibitemOpen
  \bibfield  {author} {\bibinfo {author} {\bibfnamefont {D.~S.}\ \bibnamefont
  {Bernstein}},\ }\href@noop {} {\emph {\bibinfo {title} {Matrix Mathematics:
  Theory, Facts, and Formulas}}}\ (\bibinfo  {publisher} {Princeton University
  Press},\ \bibinfo {address} {Princeton},\ \bibinfo {year} {2009})\BibitemShut
  {NoStop}%
\bibitem [{\citenamefont {Richard}\ and\ \citenamefont
  {Burden}(2005)}]{numeric}%
  \BibitemOpen
  \bibfield  {author} {\bibinfo {author} {\bibfnamefont {L.}~\bibnamefont
  {Richard}}\ and\ \bibinfo {author} {\bibfnamefont {J.}~\bibnamefont
  {Burden}},\ }\href@noop {} {\emph {\bibinfo {title} {Numerical Analysis}}}\
  (\bibinfo  {publisher} {Thomson Brooks/Cole},\ \bibinfo {address} {Belmont,
  USA},\ \bibinfo {year} {2005})\ \bibinfo {note} {douglas Faires: Numerical
  Analysis Thomson Brooks. Cole, Belmont, USA}\BibitemShut {NoStop}%
\bibitem [{\citenamefont {Wang}(2017)}]{nv1}%
  \BibitemOpen
  \bibfield  {author} {\bibinfo {author} {\bibfnamefont {Z.}~\bibnamefont
  {Wang}},\ }\emph {\bibinfo {title} {Probing Surface Spin Interaction Dynamics
  using Nitrogen-Vacancy Center Quantum Sensors with High-Fidelity
  State-Selective Transition Control}},\ \href@noop {} {Master's thesis},\
  \bibinfo  {school} {University of Waterloo} (\bibinfo {year}
  {2017})\BibitemShut {NoStop}%
\bibitem [{\citenamefont {Rossignolo}\ \emph {et~al.}(2023)\citenamefont
  {Rossignolo}, \citenamefont {Reisser}, \citenamefont {Marshall},
  \citenamefont {Rembold}, \citenamefont {Pagano}, \citenamefont {Vetter},
  \citenamefont {Said}, \citenamefont {Müller}, \citenamefont {Motzoi},
  \citenamefont {Calarco}, \citenamefont {Jelezko},\ and\ \citenamefont
  {Montangero}}]{ROSSIGNOLO2023108782}%
  \BibitemOpen
  \bibfield  {author} {\bibinfo {author} {\bibfnamefont {M.}~\bibnamefont
  {Rossignolo}}, \bibinfo {author} {\bibfnamefont {T.}~\bibnamefont {Reisser}},
  \bibinfo {author} {\bibfnamefont {A.}~\bibnamefont {Marshall}}, \bibinfo
  {author} {\bibfnamefont {P.}~\bibnamefont {Rembold}}, \bibinfo {author}
  {\bibfnamefont {A.}~\bibnamefont {Pagano}}, \bibinfo {author} {\bibfnamefont
  {P.~J.}\ \bibnamefont {Vetter}}, \bibinfo {author} {\bibfnamefont {R.~S.}\
  \bibnamefont {Said}}, \bibinfo {author} {\bibfnamefont {M.~M.}\ \bibnamefont
  {Müller}}, \bibinfo {author} {\bibfnamefont {F.}~\bibnamefont {Motzoi}},
  \bibinfo {author} {\bibfnamefont {T.}~\bibnamefont {Calarco}}, \bibinfo
  {author} {\bibfnamefont {F.}~\bibnamefont {Jelezko}},\ and\ \bibinfo {author}
  {\bibfnamefont {S.}~\bibnamefont {Montangero}},\ }\bibfield  {title}
  {\bibinfo {title} {Quocs: The quantum optimal control suite},\ }\href
  {https://doi.org/https://doi.org/10.1016/j.cpc.2023.108782} {\bibfield
  {journal} {\bibinfo  {journal} {Computer Physics Communications}\ }\textbf
  {\bibinfo {volume} {291}},\ \bibinfo {pages} {108782} (\bibinfo {year}
  {2023})}\BibitemShut {NoStop}%
\bibitem [{\citenamefont {Lim}\ \emph {et~al.}(2024)\citenamefont {Lim},
  \citenamefont {Genov}, \citenamefont {Sailer}, \citenamefont {Fahrurrachman},
  \citenamefont {Majidi}, \citenamefont {Jelezko},\ and\ \citenamefont
  {Said}}]{lim2024efficiencyoptimalcontrolnoisy}%
  \BibitemOpen
  \bibfield  {author} {\bibinfo {author} {\bibfnamefont {H.~M.}\ \bibnamefont
  {Lim}}, \bibinfo {author} {\bibfnamefont {G.~T.}\ \bibnamefont {Genov}},
  \bibinfo {author} {\bibfnamefont {R.}~\bibnamefont {Sailer}}, \bibinfo
  {author} {\bibfnamefont {A.}~\bibnamefont {Fahrurrachman}}, \bibinfo {author}
  {\bibfnamefont {M.~A.}\ \bibnamefont {Majidi}}, \bibinfo {author}
  {\bibfnamefont {F.}~\bibnamefont {Jelezko}},\ and\ \bibinfo {author}
  {\bibfnamefont {R.~S.}\ \bibnamefont {Said}},\ }\href
  {https://arxiv.org/abs/2411.05078} {\bibinfo {title} {Efficiency of optimal
  control for noisy spin qubits in diamond}} (\bibinfo {year} {2024}),\ \Eprint
  {https://arxiv.org/abs/2411.05078} {arXiv:2411.05078 [quant-ph]} \BibitemShut
  {NoStop}%
\bibitem [{\citenamefont {Lambert}\ \emph {et~al.}(2024)\citenamefont
  {Lambert}, \citenamefont {Giguère}, \citenamefont {Menczel}, \citenamefont
  {Li}, \citenamefont {Hopf}, \citenamefont {Suárez}, \citenamefont {Gali},
  \citenamefont {Lishman}, \citenamefont {Gadhvi}, \citenamefont {Agarwal},
  \citenamefont {Galicia}, \citenamefont {Shammah}, \citenamefont {Nation},
  \citenamefont {Johansson}, \citenamefont {Ahmed}, \citenamefont {Cross},
  \citenamefont {Pitchford},\ and\ \citenamefont {Nori}}]{qutip}%
  \BibitemOpen
  \bibfield  {author} {\bibinfo {author} {\bibfnamefont {N.}~\bibnamefont
  {Lambert}}, \bibinfo {author} {\bibfnamefont {E.}~\bibnamefont {Giguère}},
  \bibinfo {author} {\bibfnamefont {P.}~\bibnamefont {Menczel}}, \bibinfo
  {author} {\bibfnamefont {B.}~\bibnamefont {Li}}, \bibinfo {author}
  {\bibfnamefont {P.}~\bibnamefont {Hopf}}, \bibinfo {author} {\bibfnamefont
  {G.}~\bibnamefont {Suárez}}, \bibinfo {author} {\bibfnamefont
  {M.}~\bibnamefont {Gali}}, \bibinfo {author} {\bibfnamefont {J.}~\bibnamefont
  {Lishman}}, \bibinfo {author} {\bibfnamefont {R.}~\bibnamefont {Gadhvi}},
  \bibinfo {author} {\bibfnamefont {R.}~\bibnamefont {Agarwal}}, \bibinfo
  {author} {\bibfnamefont {A.}~\bibnamefont {Galicia}}, \bibinfo {author}
  {\bibfnamefont {N.}~\bibnamefont {Shammah}}, \bibinfo {author} {\bibfnamefont
  {P.}~\bibnamefont {Nation}}, \bibinfo {author} {\bibfnamefont {J.~R.}\
  \bibnamefont {Johansson}}, \bibinfo {author} {\bibfnamefont {S.}~\bibnamefont
  {Ahmed}}, \bibinfo {author} {\bibfnamefont {S.}~\bibnamefont {Cross}},
  \bibinfo {author} {\bibfnamefont {A.}~\bibnamefont {Pitchford}},\ and\
  \bibinfo {author} {\bibfnamefont {F.}~\bibnamefont {Nori}},\ }\bibfield
  {title} {\bibinfo {title} {Qutip 5: The quantum toolbox in python},\ }\href
  {https://arxiv.org/abs/2412.04705} {\  (\bibinfo {year} {2024})},\ \Eprint
  {https://arxiv.org/abs/2412.04705} {arXiv:2412.04705 [quant-ph]} \BibitemShut
  {NoStop}%
\bibitem [{\citenamefont {Tiesinga}\ \emph {et~al.}(2021)\citenamefont
  {Tiesinga}, \citenamefont {Mohr}, \citenamefont {Newell},\ and\ \citenamefont
  {Taylor}}]{egyro}%
  \BibitemOpen
  \bibfield  {author} {\bibinfo {author} {\bibfnamefont {E.}~\bibnamefont
  {Tiesinga}}, \bibinfo {author} {\bibfnamefont {P.~J.}\ \bibnamefont {Mohr}},
  \bibinfo {author} {\bibfnamefont {D.~B.}\ \bibnamefont {Newell}},\ and\
  \bibinfo {author} {\bibfnamefont {B.~N.}\ \bibnamefont {Taylor}},\ }\bibfield
   {title} {\bibinfo {title} {Codata recommended values of the fundamental
  physical constants: 2018},\ }\href
  {https://doi.org/10.1103/RevModPhys.93.025010} {\bibfield  {journal}
  {\bibinfo  {journal} {Rev. Mod. Phys.}\ }\textbf {\bibinfo {volume} {93}},\
  \bibinfo {pages} {025010} (\bibinfo {year} {2021})}\BibitemShut {NoStop}%
\bibitem [{\citenamefont {Doherty}\ \emph {et~al.}(2013)\citenamefont
  {Doherty}, \citenamefont {Manson}, \citenamefont {Delaney}, \citenamefont
  {Jelezko}, \citenamefont {Wrachtrup},\ and\ \citenamefont
  {Hollenberg}}]{nv3}%
  \BibitemOpen
  \bibfield  {author} {\bibinfo {author} {\bibfnamefont {M.~W.}\ \bibnamefont
  {Doherty}}, \bibinfo {author} {\bibfnamefont {N.~B.}\ \bibnamefont {Manson}},
  \bibinfo {author} {\bibfnamefont {P.}~\bibnamefont {Delaney}}, \bibinfo
  {author} {\bibfnamefont {F.}~\bibnamefont {Jelezko}}, \bibinfo {author}
  {\bibfnamefont {J.}~\bibnamefont {Wrachtrup}},\ and\ \bibinfo {author}
  {\bibfnamefont {L.~C.}\ \bibnamefont {Hollenberg}},\ }\bibfield  {title}
  {\bibinfo {title} {The nitrogen-vacancy colour centre in diamond},\
  }\href@noop {} {\bibfield  {journal} {\bibinfo  {journal} {Physics Reports}\
  }\textbf {\bibinfo {volume} {528}},\ \bibinfo {pages} {1} (\bibinfo {year}
  {2013})}\BibitemShut {NoStop}%
\bibitem [{\citenamefont {Haruyama}\ \emph {et~al.}(2019)\citenamefont
  {Haruyama}, \citenamefont {Onoda}, \citenamefont {Higuchi} \emph
  {et~al.}}]{triplenv}%
  \BibitemOpen
  \bibfield  {author} {\bibinfo {author} {\bibfnamefont {M.}~\bibnamefont
  {Haruyama}}, \bibinfo {author} {\bibfnamefont {S.}~\bibnamefont {Onoda}},
  \bibinfo {author} {\bibfnamefont {T.}~\bibnamefont {Higuchi}}, \emph
  {et~al.},\ }\bibfield  {title} {\bibinfo {title} {Triple nitrogen-vacancy
  centre fabrication by c$_5$n$_4$h$_n$ ion implantation},\ }\href
  {https://doi.org/10.1038/s41467-019-10529-x} {\bibfield  {journal} {\bibinfo
  {journal} {Nature Communications}\ }\textbf {\bibinfo {volume} {10}},\
  \bibinfo {pages} {2664} (\bibinfo {year} {2019})}\BibitemShut {NoStop}%
\bibitem [{plo()}]{plotdata}%
  \BibitemOpen
  \href@noop {} {}\bibinfo {howpublished} {{Zenodo or github link after
  review}}\BibitemShut {NoStop}%
\end{thebibliography}
\end{document}